\documentclass[aps,prd,twocolumn,
superscriptaddress,
floatfix,
showpacs,showkeys,preprintnumbers]{revtex4}
\usepackage{pdfsync}
\usepackage{mathptmx}
\usepackage[utf8]{inputenc}
\usepackage[T1]{fontenc}
\usepackage{slashed}
\usepackage{times}
\usepackage{amssymb,amsfonts,amsmath,amsthm}
\usepackage{dsfont,bbm}
\usepackage{dcolumn}
\usepackage{epsf}
\usepackage{graphicx}
\usepackage[caption=false]{subfig}
\usepackage{dsfont}
\usepackage{slashed}
\usepackage[active]{srcltx}
\usepackage[usenames]{color}

\begin{document}

\title{Gravitational form factors within light-cone sum rules at leading order}

\author{I.~V.~Anikin}
\email{anikin@theor.jinr.ru}
\affiliation{Bogoliubov Laboratory of Theoretical Physics, JINR,
             141980 Dubna, Russia}

\begin{abstract}
We develop an approach based on the light-cone sum rules at the leading order of $\alpha_S$ to calculate
the gravitational form factors $\mathds{A}(t)$ and $\mathds{B}(t)$ for the valence quark combinations in nucleon.
Within the proposed model, the predictions for the gravitational form factor $\mathds{D}(t)$ ($D$-term contributions) have been presented.
Comparison with the experimental data and with the results of different models is discussed.
\end{abstract}
\pacs{12.38.-t, 14.20.Dh; 13.40.Gp}
\keywords{Gravitational form factors, QCD, Electromagnetic form factors, Light-cone sum rules, $D$-term}
\date{\today}
\maketitle

\section{Introduction}
\label{Intro}

It is well-known that the hadron matrix element of energy-momentum tensor (EMT) can provide information
on fundamental characteristics of particles such as mass and spin \cite{Polyakov:2002yz, Polyakov:2018zvc, Polyakov:2018exb, Belitsky:2005qn, Diehl:2003ny}.
Especially, efforts have been made to establish the relations between the gravitational form factors $\mathds{A}(t)$ and $\mathds{B}(t)$
which parameterize
the hadron matrix element of EMT and the Mellin $x$-moment of generalized parton distributions
$H(x, \xi, t)$ and $E(x, \xi, t)$ also known as Ji's sum rules.
There is an opinion that such relations can help us to make a progress in understanding the hadron spin problem
(see, for example, comprehensive reviews \cite{Belitsky:2005qn, Diehl:2003ny}).
Besides, one of the EMT form factors has been related to the so-called $D$-term \cite{Kivel:2000fg} which is considered
as the last unknown fundamental hadron characteristic determining the spatial deformations as well as
defining the mechanical properties of hadrons \cite{Polyakov:2002yz, Polyakov:2018zvc, Polyakov:2018exb}.
The analogy of $D$-term with the vacuum cosmological constant has been observed in \cite{Teryaev:2013qba}.
Also, the $D$-term has been calculated using dispersion relations giving the good agreement with
the chiral quark-soliton model \cite{Pasquini:2014vua}.
Recently, the results of \cite{Polyakov:2018zvc} have been extended to the different
frames where the nucleon has the non-vanishing average momentum \cite{Lorce:2018egm}.

The energy-momentum tensor plays the role in the interplay between the gravitation as an external field
and the matter fields
in the similar manner as the gauge field (photons, gluons) interacts with fermions or other particles by
means of the corresponding electromagnetic current. In this connection, we shall adopt
the technique of light-cone sum rules (LCSRs) developed for the different nucleon electromagnetic form factors in
\cite{Anikin:2013aka, Anikin:2015ita, Anikin:2016teg}.

In the present paper, we develop the light-cone sum rules for
the purpose of computing the nucleon gravitational form factors.
In \cite{Broniowski:2008hx, Kumano:2017lhr},
the pion gravitational form factors have been studied using the hadron tomography and effective chiral quark model.
The obvious preponderance of LCSRs is that it provides a possibility to calculate the soft contributions
to the different form factors as an
expansion in terms of hadron distribution amplitudes (DAs) of increasing twist with the help of
dispersion relations and quark-hadron duality \cite{Balitsky:1986st, Balitsky:1989ry, Braun:2006hz, Braun:2001tj}.
Indeed, within the LCSRs formalism, the soft contributions to the form factors can be calculated in terms of the same DAs that
enter the factorization theorem and/or pQCD calculations. More importantly, it has been proven that
there is no double counting (see, for example, \cite{Balitsky:1986st}). Thus, the LCSRs provide one
with the most direct relation of the hadron form factors and
DAs that is available at present, without requiring other nonperturbative
parameters.

In the frame of the approach that we develop based on the light-cone sum rules
at the leading order of $\alpha_S$,
we compute the gravitational form factors $\mathds{A}(t)$ and $\mathds{B}(t)$
and estimate the gravitational form factor $\mathds{D}(t)$.
We emphasize that the gravitational form factors $\mathds{A}(t)$ and $\mathds{B}(t)$ can be
calculated directly with the help of the sum rules extracted from the plus-plus light-cone projection of EMT
for sufficiently large Euclidian $t\gtrsim 1\, \text{GeV}^2$.
The plus-plus light-cone projection of EMT can be associated with the plus light-cone
projection of electromagnetic current.
Unfortunately, as explained in this paper, the full information on the gravitation form factor
$\mathds{D}(t)$  ($D$-term form factor) cannot be obtained within the LCSRs due to
the lack of the sum rules that use the minus and perpendicular light-cone projections of electromagnetic current.
Instead, for the estimation of $\mathds{D}(t)$, we study the valence quark contributions in nucleon that stem
from the leading collinear twist-$2$ combination of the electromagnetic current.

In order to approach the small $t$ region, where the LCSRs approach is actually useless,
we first approximate the form factors derived for the large $t$ region by
the appropriate multipole functions and, then we do the analytical continuations of fited form factors to
the small $t$ region.  With the calculated form factors, we present the result for
the Mellin $x$-moment of GPDs combination $H(x, \xi, t) + E(x, \xi, t)$.
Our final predictions demonstrate reasonably good agreements with the first experimental data \cite{Hagler:2007xi, Burkert:2018bqq},
the chiral quark-soliton and Skyrme model results for $D$-term contributions \cite{Polyakov:2018zvc}.

The presentation is organized as follows. Sect.~\ref{Subsec:emt-tw} contains
the necessary definitions and explanations of our notations.
Sect.~\ref{Sec:GFF-LO} includes the description of the general structure of LCSRs approach, the leading-order
sum rules and the results concerning the gravitational form factors and their applications for the description of the
so-called mechanical properties of nucleon.
Sect.~\ref{Sec:Conclusions} is reserved for conclusions.
In App.~\ref{App} we give the introductory material about the geometrical and collinear twists.

\section{Energy-momentum tensor}
\label{Subsec:emt-tw}

Let us introduce the Sudakov expansion of an arbitrary vector as
\begin{eqnarray}
a^\mu = a^+ p^\mu + a^- n^\mu + a^\mu_\perp\equiv a^{\mu,+} + a^{\mu,-} + a^\mu_\perp
\end{eqnarray}
with
\begin{eqnarray}
p^2=n^2=0, \quad p\cdot n =1.
\end{eqnarray}
Throughout the paper, for the sake of simplicity, we assume all Lorentz indices to be contravariant ones irrespective of
its positions in our formulae unless it is specified otherwise, {\it i.e.}
\begin{eqnarray}
A_\alpha B_\alpha  \equiv A_\alpha B^\alpha=A^+B^- + A^-B^+ - \vec{\bf A}_\perp\vec{\bf B}_\perp.
\end{eqnarray}

For the quark contribution, the Belinfant\'{e} improved energy-momentum tensor
is nothing but the local geometrical twist-$2$ operator which reads
\begin{eqnarray}
\label{emt-1}
\Theta^{\mu\nu}_q(0)= \frac{i}{2} {\cal R}^{\mu\nu}_{\tau=2}
\end{eqnarray}
and it can be expressed through the non-local operator as (cf. \cite{Balitsky:1987bk})
\begin{eqnarray}
\label{emt-2}
-2i\,\Theta^{\mu\nu}_q(0)&=&\lim_{y\to 0}\frac{\partial}{\partial y_\nu}\, \int\limits_{0}^{1}du\,\frac{\partial}{\partial y_\mu}
\big[ \bar\psi(0)\, \hat y \,[0\,;\,uy]_A\, \psi(uy)  \big]
\nonumber\\
&-& \text{(trace)},
\end{eqnarray}
where $y=(y^+,y^-,\vec{\bf y}_\perp)$. From now on, we do not show the trace subtractions.
In Appendix~\ref{App}, we briefly provide the necessary descriptions regarding the collinear and geometrical twists.

In addition, for our further purposes, we also introduce the local geometrical twist-2 operator defined as
\begin{eqnarray}
\label{tw-2-2}
&&\widetilde{\cal R}^{\mu\nu}_{\tau=2}=\lim_{y\to 0}\frac{\partial}{\partial y_\nu}\, \int\limits_{0}^{1}du\,\frac{\partial}{\partial y_\mu}
\big[ \bar\psi(0)\, y_\alpha\gamma^{\alpha,+} \,[0\,;\,uy]_A\, \psi(uy)  \big]
\nonumber\\
&&=\frac{1}{2} \Big( \bar\psi(0)\, \gamma^{\mu,+}\vec{\cal D}^{\nu}\,\psi(0) +  \bar\psi(0)\, \gamma^{\nu,+}\vec{\cal D}^{\mu}\,\psi(0) \Big)
\nonumber\\
&&=\frac{1}{2} \Big( g^{\mu -} \bar\psi(0)\, \gamma^{+}\vec{\cal D}^{\nu}\,\psi(0) + g^{\nu -} \bar\psi(0)\, \gamma^{+}\vec{\cal D}^{\mu}\,\psi(0) \Big)
\end{eqnarray}
which differs from the operator ${\cal R}^{\mu\nu}_{\tau=2}$. However, it is important to note that
\begin{eqnarray}
\label{R-tildeR}
\widetilde{\cal R}^{+ +}_{\tau=2} = {\cal R}^{+ +}_{\tau=2}.
\end{eqnarray}
The reason for the introduction of $\widetilde{\cal R}^{\mu\nu}_{\tau=2}$ is the following:
it is well-known \cite{Braun:2006hz} that for the electromagnetic form factors the LCSRs can be established self-consistently only
for the plus light-cone projection of electromagnetic current, $J_{em}^+(0)$, which corresponds to the twist-$2$ operator combination of
the current.
Having kept only the plus light-cone projection of the non-local operator,
we are able to develop LCSRs for the gravitational form factors by means of
appropriate adoption of our preceding calculations implemented for
the electromagnetic form factors \cite{Anikin:2013aka} (see eqns.~(\ref{emt-2}) and (\ref{tw-2-2})).
As a result, we are limited by the bilinear quark combinations with the spin projection $s_a=+1$,
{\it  i.e.} we deal with the collinear twist-$2$ quark combination $[\bar\psi_+\, \psi_+]$ of ${\cal R}^{\mu\nu}_{\tau=2}$.
For example, excluding the trivial case of the plus-plus projection presented in eqn.~(\ref{R-tildeR}),
we consider ${\cal R}^{+-}_{\tau=2}$ which is
\begin{eqnarray}
\label{pmR}
{\cal R}^{+-}_{\tau=2}=&&
\frac{1}{2} \Big( \bar\psi(0)\, \gamma^{+}\vec{\cal D}^{-}\,\psi(0) +  \bar\psi(0)\, \gamma^{-}\vec{\cal D}^{+}\,\psi(0) \Big)
\nonumber\\
=&&\widetilde{\cal R}^{+-}_{\tau=2}  + \frac{1}{2} \bar\psi(0)\, \gamma^{-}\vec{\cal D}^{+}\,\psi(0).
\end{eqnarray}
Here, the first term with $[\bar\psi_+\, \psi_+]$-combination is traded for $\widetilde{\cal R}^{+-}_{\tau=2}$ while
the second term with $[\bar\psi_-\, \psi_-]-$combination is kept intact and it is beyond the direct computations within our approach.

Notice that the $[\bar\psi_-\, \psi_-]-$ and $[\bar\psi_+\, \psi_-]-$contributions to $\Theta^{+-}$ and $\Theta^{+\perp}$
remain unavailable for the explicit LCSRs calculations,
because they are given by ``bad'' projections $J_{em}^-(0)$ and $J_{em}^\perp(0)$.
Indeed, in this case
we are forced to deal with the amplitudes generated by the vacuum-nucleon matrix element of T-product involving
the Ioffe interpolation current $\eta$ (see eqn.~(\ref{Ioffe})) and
operator $\bar\psi\, \gamma^{-(\perp)}\vec{\cal D}^{+}\,\psi$, {\it i.e.}
$\langle 0| T \eta(0) [\bar\psi(x)\, \gamma^{-(\perp)}\vec{\cal D}^{+}\,\psi(x)]|P\rangle$,
which, in turn, are related to the useless amplitude with
$\langle 0| T \eta(0) J_{em}^{-(\perp)}(x) |P\rangle$.

However, we can still make several estimations for this kind of contributions.

\section{Gravitational Form Factors within the LO LCSR}
\label{Sec:GFF-LO}

Before we move on to the detailed analysis, let us draw the main stages of our approach based on the leading order LCSRs.
First, we begin with the quark contribution to the hadron matrix element of the Belinfant\'{e} improved energy-momentum tensor operator
which is (see, for instance, \cite{Polyakov:2018exb})
\begin{eqnarray}
\label{EMT-me-1}
&&\langle P^\prime | \Theta^{(q)}_{\mu\nu}(0) | P \rangle =
\bar N(P^\prime) \Big[ \mathds{A}(t)\frac{\overline{P}_\mu\overline{P}_\nu}{m_N} +
i\mathds{J}(t)\frac{\overline{P}_{\left\{\mu\right.}\sigma_{\left.\nu\right\}\Delta}}{m_N} +
\nonumber\\
&&
\mathds{D}(t)\frac{\Delta_\mu\Delta_\nu -g_{\mu\nu} \Delta^2}{4m_N} + g_{\mu\nu} m_N \overline{\mathds{C}}(t)
\Big] N(P),
\end{eqnarray}
where
\begin{eqnarray}
&&\mathds{J}(t)=\frac{1}{2} \Big( \mathds{A}(t) + \mathds{B}(t)\Big),\quad
a_{\left\{\mu\right.}b_{\left.\nu\right\}} = \frac{1}{2} \big( a_\mu b_\nu + a_\nu b_\mu\big), \quad
\nonumber\\
&&
\overline{P}=\frac{1}{2}\big( P^\prime + P\big), \quad \Delta=P^\prime - P, \quad \Delta^2=-t.
\end{eqnarray}
Notice that the Belinfant\'{e} improved energy-momentum or angular momentum tensor includes the contribution from the spin momentum tensor.

As usual, the hadron momenta can be expanded over the light-cone basis as \cite{Anikin:2013aka}
\begin{eqnarray}
P=p+\frac{m^2_N}{2}n, \quad \Delta=P\cdot \Delta\, n + \Delta_\perp.
\end{eqnarray}

As shown below, the projection ${\cal R}^{++}_{\tau=2}$ is enough to calculate only the form factors $\mathds{A}(t)$ and $\mathds{B}(t)$.

Regarding the form factor $\overline{\mathds{C}}(t)$, based on the QCD equations of motion
it can be expressed through the nucleon matrix elements of the quark-gluon operator
(for the quark contribution) or the gluon-gluon operator (for the gluon contribution) \cite{Tanaka:2018wea}.
Therefore, the calculation of $\overline{\mathds{C}}(t)$ is not available at the leading order.
Moreover, this form factor meets the condition $\sum_a \overline{\mathds{C}^a}(t)=0$ which shows that
the total (quark and gluon) EMT is conserved.
So, at the moment, the consideration of $\overline{\mathds{C}}(t)$ is not presented in the
present paper and is postponed to the forthcoming work.

To calculate the form factor $\mathds{D}(t)$ one needs a projection such as ${\cal R}^{+-}_{\tau=2}$.
In the present paper, our exact computations are restricted by the consideration of $[\bar\psi_+ \, \psi_+]-$combinations.
Hence, we are dealing with the corresponding
projections of $\widetilde{\cal R}^{\mu\nu}_{\tau=2}$ rather than the projections of ${\cal R}^{\mu\nu}_{\tau=2}$.
Regarding the $[\bar\psi_- \, \psi_-]-$combinations, we propose a reliable recipe for estimations of these contributions,
see the subsection~\ref{Subsec:plus-minus-proj}.

In analogy with the nucleon electromagnetic form factors,
we define the amplitude which corresponds to the hadron matrix element of the energy-momentum tensor as
\begin{eqnarray}
\label{gravi-amp-1}
&&\hspace{-0.4cm}T_{\mu\nu}(P,\Delta)=\lim_{y\to 0}\frac{\partial}{\partial y_\nu}\, \int\limits_{0}^{1}du\,\frac{\partial}{\partial w_\mu}
\int(d^4z) e^{-i\Delta\cdot z} \langle 0| T \eta(0)
\nonumber\\
&&\hspace{-0.4cm}
\times\big[ \bar\psi(w+z)\, w^\alpha\gamma^{\alpha,+}\,
[w+z\,;\,-w+z]_A\, \psi(-w+z) \big]| P\rangle
\end{eqnarray}
where $w=uy$. $\eta(0)$ stands for the Ioffe interpolation current defined as
\begin{eqnarray}
\label{Ioffe}
\eta(0)=\varepsilon^{ijk}\big[ u^i(0) C \gamma_\alpha u^j(0)\big] \gamma_5 \gamma_\alpha d^k(0).
\end{eqnarray}
Below, the colour indices $i,j,k,..$ are omitted.

As mentioned above, we adhere to the computation procedure applied first to the case of electromagnetic form factors \cite{Anikin:2013aka}.
After the replacement $z\to z+w$, the amplitude (\ref{gravi-amp-1}) takes the general form of
\begin{eqnarray}
\label{gravi-amp-2}
&&\hspace{-0.2cm}T_{\mu\nu}(P,\Delta)={\cal P}^{(\eta)} \int(d^4 z) e^{-i\Delta\cdot (z+w)} \int [dx_i] e^{-ix_iP\cdot z}
\times\\
&&\hspace{-0.2cm}
\Big\{ \mathbb{V}_1(x) \big( \hat P C\big)\, \gamma_5 N(P) + ....\Big\}
\otimes \Big[ \int (d^4 k) e^{+ik\cdot(z+2w)}\, S(k) \,\bar\mu \hat n\Big]
\nonumber
\end{eqnarray}
where  $w^\alpha\gamma^{\alpha,+}=\bar\mu \hat n$,
${\cal P}^{(\eta)}$ implies the Ioffe interpolation current projection operator
and different $\mathbb{V}_i(x)$ denote the corresponding DAs.
For the sake of shortness, we here use the shorthand notation:
\begin{eqnarray}
\label{amp-sh-note}
{\cal P}^{(\eta)}
\Big\{ \mathbb{V}_1(x) \big( \hat P C\big)\, \gamma_5 N(P) + ....\Big\}
\otimes \big[ \,S(k) \,\hat n\big],
\end{eqnarray}
which, in the case of $d$-quark contribution,  is equal to
\begin{eqnarray}
\label{amp-sh-note-d}
\mathbb{V}_1(x) \text{tr}\big[C \gamma_\alpha \hat P C\big]\, \gamma_5 \gamma_\alpha \,S(k)\hat n \,\gamma_5 N(P) + ....
\end{eqnarray}
The extension to the other flavours is straightforward.

Notice that our DAs are, generally speaking,
scheme- and scale-dependent. In calculations of any
physical observables this dependence is cancelled by the corresponding
dependence of the coefficient functions. Although the $\mu^2$-dependence of DAs
is not shown, we always keep in mind this dependence.
Moreover, our DAs can be expanded in the set of orthogonal polynomials
${\cal P}_{nk}(x_i)$ defined as eigenfunctions of the corresponding
one-loop evolution equation. In our previous study \cite{Anikin:2013aka},
the different parameter dependences of DAs have been thoroughly investigated.
We extend this analysis to our present work.

A little algebra, using integration by parts, leads to the following result
\begin{eqnarray}
\label{gravi-amp-3}
&&T_{\mu\nu}(P,\Delta)=
\\
&&
\frac{i}{2}\,\frac{(-i)}{4}{\cal P}^{(\eta)} \int {\cal D}x_i
\Big(\frac{\partial}{\partial x_1} \big[ 2x_1 P + \Delta \big]_{\mu} \big[ 2x_1 P + \Delta \big]_{\nu} \Big)
\nonumber\\
&&\times
\Big\{ \mathbb{V}_1(x) \big( \hat P C\big)\, \gamma_5 N(P) + ....\Big\}
\otimes \Big[ S(x_1P + \Delta) \,\bar\mu \hat n\Big].
\nonumber
\end{eqnarray}
To extend the preceding calculations performed for
electromagnetic form factors to the case of gravitational form factors
we have to weight the electromagnetic form factors with the certain tensor structure, see (\ref{gravi-amp-3}).
Indeed, in the schematic form, the correspondence reads
\begin{eqnarray}
\label{Adopt-1}
&&\Big\{ \text{Grav.FFs}\Big\}_{\mu\nu} =
\\
&&
\Big( 2p_{\mu} \big[ 2x_i P + \Delta \big]_{\nu} + (\mu\leftrightarrow\nu) \Big)
\otimes \Big\{ \text{Elec.Mag.FFs}\Big\}
\nonumber
\end{eqnarray}
The exact expression for the amplitude for $d$-quark contribution is given by
\begin{eqnarray}
\label{Adopt-2}
&&T^{(d)}_{\mu\nu}(P,\Delta) =
\\
&&
\frac{1}{8} \int {\cal D}x_i
\Big( 2p_{\mu} \big[ 2x_1 P + \Delta \big]_{\nu} + (\mu\leftrightarrow\nu) \Big)
T^{(d)\,+}_{em}(x_i; P,\Delta),
\nonumber
\end{eqnarray}
where
\begin{eqnarray}
\label{Adopt-2-2}
&&T^{(q)\,+}_{em}(x_i; P,\Delta) =
\\
&&
 \Big\{
m_N\, {\cal A}^{+\,(q)}_{em}(x_i; \Delta^2, P^{\prime\, 2}) +
\hat\Delta_\perp \,{\cal B}^{+\,(q)}_{em}(x_i; \Delta^2, P^{\prime\, 2})\Big\} N^+(P),
\nonumber\\
&&N^\pm(P)= \Lambda^\pm N(P), \quad \Lambda^+ = \frac{\hat p\, \hat n}{2 pn}, \quad 
\Lambda^- = \frac{\hat n\,\hat p}{2 pn}.
\nonumber
\end{eqnarray}
and ${\cal A}^{+\,(q)}_{em}$, ${\cal B}^{+\,(q)}_{em}$ have been taken from \cite{Anikin:2013aka}
\footnote{
For the sake of convenience, we indicate the following correspondence with
the relevant formulas from Ref.~\cite{Anikin:2013aka}:
our $Q^2{\cal A}^{+\,(d,u)}_{em}$ is given by the integrand of $Q^2 \mathcal{A}^{\mathrm{LO}}_{d,u}$
from Eqn.~(44) of \cite{Anikin:2013aka}, while our
$Q^2{\cal B}^{+\,(d,u)}_{em}$ is given by the integrand of $Q^2 \mathcal{B}^{\mathrm{LO}}_{d,u}$
from Eqn.~(45) of \cite{Anikin:2013aka}.
}.

From eqn.~(\ref{Adopt-2}), we conclude that the tensor structure of the amplitude (\ref{gravi-amp-1})
is entirely determined by the tensor
\begin{eqnarray}
2p_{\mu} \big[ 2x_i P + \Delta \big]_{\nu} + (\mu\leftrightarrow\nu).
\end{eqnarray}

Since the Ioffe current $\eta(0)$ has been used as the interpolation current, we focus on the
valence quark contributions to nucleon. Notice that all amplitudes and form factors should be understood as the objects
where the summation over $u$- and $d$-flavours have been implemented:
$T_{\mu\nu}=T^{(u)}_{\mu\nu}+T^{(d)}_{\mu\nu}$ etc.

\subsection{Plus-plus light-cone projections of the amplitude: form factors $ \mathds{A}(t)$ and $ \mathds{B}(t)$}
\label{Subsec:plus-plus-proj}

We are now in a position to discuss the LCSRs which stem from the different light-cone projections.
To begin with, we dwell on the plus-plus light-cone projection of the amplitude which is given by
\begin{eqnarray}
&&n_\mu\,n_\nu \, T^{\mu\nu}(P,\Delta)=T^{++}(P,\Delta)=
\nonumber\\
&&
\Big[ m_N\, \mathcal{A}^{++}(P,\Delta) + \hat\Delta_\perp \mathcal{B}^{++}(P,\Delta)\Big] N^+(P).
\end{eqnarray}
Making use of the Borel transforms one obtains the sum rules
\begin{eqnarray}
\label{A-B-ff}
&&\mathds{A}(t) = \frac{1}{2} \int \hat d\mu(s)\Big\{\mathcal{A}^{++}_{{\rm QCD}}(\Delta^2,s) \Big\}
\\
&& \mathds{B}(t) = -\int \hat d\mu(s)\Big\{\mathcal{B}^{++}_{{\rm QCD}}(\Delta^2,s) \Big\}.
\end{eqnarray}
where, for the sake of shortness, we introduce the convenient notation
\begin{eqnarray}
\label{DR-short}
\int \hat d\mu(s)\big\{\mathcal{F} \big\}
=\frac{1}{\lambda_1\pi}\int_{0}^{s_0} ds\, e^{(m_N^2-s)/M^2}\, \text{Im} \big\{\mathcal{F}(s,t) \big\}.
\end{eqnarray}
In eqn.~(\ref{DR-short}), $s_0\simeq (1.5$~GeV$)^2$ implies the interval of duality (also called continuum threshold)
and $\lambda_1$ parameterizes the vacuum-nucleon matrix element of the interpolation current $\eta$,
$\langle 0| \eta(0)|P\rangle =\lambda_1 m_N N(P)$.

In contrast to the other gravitational form factors,
these form factors can totally be calculated in QCD with a help of the factorization theorem
for sufficiently large Euclidian $t\gtrsim 1\, \text{GeV}^2$.
The leading order expressions for $\mathcal{A}^{++}$ and  $\mathcal{B}^{++}$
are available from \cite{Anikin:2013aka} provided the fraction integrations are weighted with the corresponding $x_i$-variables.
That is, we should add $x_3$ or $x_2$ in the integration measures for $d$- or $u$-quark contributions of $\mathcal{A}$-
and $\mathcal{B}$-amplitudes.

In \cite{Anikin:2013aka}, for consistency with our next-to-leading calculation we rewrite the leading order results in
the form where all kinematic factors have been expanded in powers of $m^2_N/t$.
Here, we also follow this method and keep all corrections $O(m^2_N/t)$ while the terms $O(m^4_N/t^2)$
have been neglected. This is consistent with taking into account the contributions of
twist-three, -four, -five (and, partially, twist-six) in the operator product expansion.
Notice that the most higher twist contributions are presented by the Wandzura-Wilczek
approximation, except the twist-four DAs where the ``genuine'' twist-four has been involved.
The genuine twist-five and twist-six distribution amplitudes are not known and,
in our analysis, these contributions are neglected. This is consistent
with neglecting four-particle nucleon DA terms with the additional gluon.

Therefore, according to this strategy, the gravitational form factors $\mathds{A}(t)$ and $\mathds{B}(t)$
are not reachable for the region of small $t$ within LCSRs approach.
However, we can do approximate fits for these form factors in the region of large $t$ and, then,
perform an analytical continuation of the obtained fitting functions to the region of small $t$.

\begin{figure}[ht]
\vspace{0.3cm}
\includegraphics[width=0.35\textwidth]{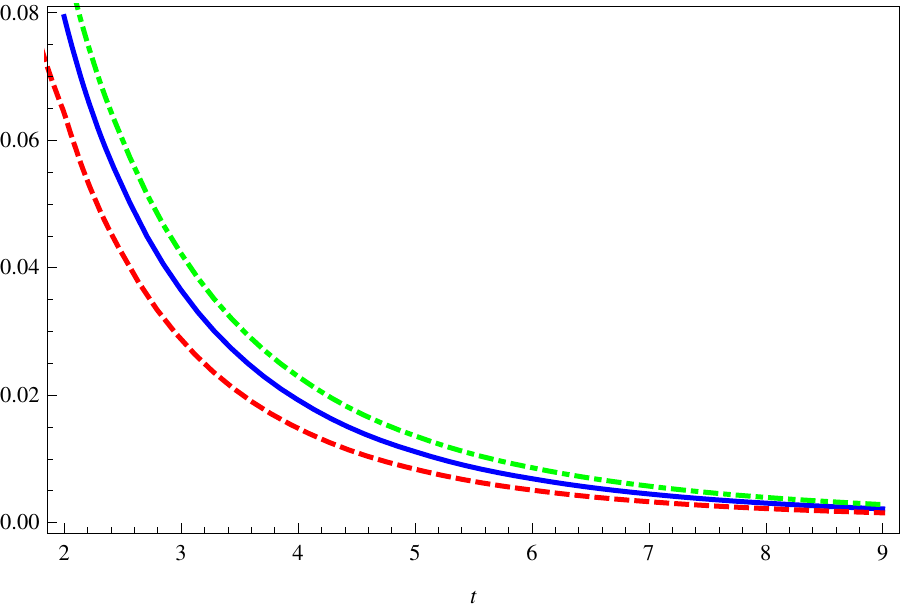}
  \caption{The gravitational form factor $\mathds{A}(t)=\mathds{A}^{(u)}(t)+\mathds{A}^{(d)}(t)$
for the region of large $t$.
  Notations: the solid blue line corresponds to ABO2; the dashed red line corresponds to ABO1; the dot-dashed green
  line corresponds to BLW \cite{Anikin:2013aka}.}
\label{Fig-1}
\end{figure}

\begin{figure}[ht]
\vspace{0.3cm}
\includegraphics[width=0.35\textwidth]{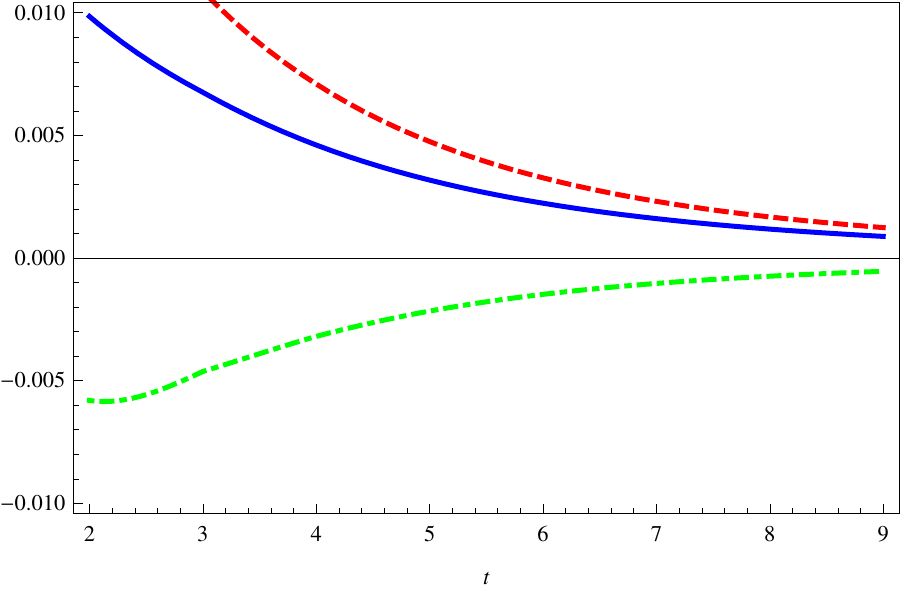}
  \caption{The gravitational form factor $-\mathds{B}(t)=-\mathds{B}^{(u)}(t)-\mathds{B}^{(d)}(t)$
 for the region of large $t$.
  Notations: the solid blue line corresponds to ABO2; the dashed red line corresponds to ABO1; the dot-dashed green
  line corresponds to BLW \cite{Anikin:2013aka}.}
\label{Fig-2}
\end{figure}

The computation results for $\mathds{A}(t)$ and $\mathds{B}(t)$ are presented in Figs.~\ref{Fig-1} and \ref{Fig-2}, respectively.
It turns out that the gravitational form factor $\mathds{A}(t)$ can be sufficiently described by the
multipole function defined as
\begin{eqnarray}
\label{A-fit}
&&
\mathds{A}^{\text{fit}}(t) = \frac{\kappa}{(1+a\, t)^b},\,
\nonumber\\
&&
\kappa=1.01\pm 0.13, \quad a=0.7\pm 0.05\,\text{GeV}^{-2},
\nonumber\\
&&
 b= 2.95\pm 0.05, \quad \mathds{A}^{\text{fit}\,\prime}(0)=-2.1,
\end{eqnarray}
while the gravitational form factor $\mathds{B}(t)$ can be approximated with
the function given by
\begin{eqnarray}
\label{B-fit}
&&\mathds{B}^{\text{fit}}(t) =
\frac{k\, t}{(1+ c\, t)^d},
\nonumber\\
&&
k=0.073\pm 0.003, \quad c=0.45\pm 0.02\,\text{GeV}^{-2},
\nonumber\\
&&
d=4.1\pm 0.2
\end{eqnarray}
both for ABO2-parametrization of \cite{Anikin:2013aka}.

As in Ref.~\cite{Anikin:2013aka}, main nonperturbative input in the LCSR calculation of form factors is provided by
normalization constants and shape parameters of nucleon DAs:
$\{\varphi_{10}, \varphi_{11}, \varphi_{20}, \varphi_{21}, \varphi_{22}\}$ -- for twist-$3$ and
$\{\eta_{10}, \eta_{11}\}$ -- for twist-$4$.
The nucleon coupling to the (Ioffe) interpolation current $\lambda_1$ simultaneously
determines the normalization of twist-four DAs
so that the LCSRs effectively only involve the ratio of twist-three and twist-four couplings,
$f_N/\lambda_1$. All parameters are rescaled to $\mu^2=2$~GeV$^2$ using one-loop anomalous
dimensions. The other parameters that enter LCSRs are the interval of duality (continuum threshold) $s_0$ and Borel parameter $M^2$.
We do separate the results for $M^2=1.5$~GeV$^2$ and  $M^2=2$~GeV$^2$ that are referred in what follows as ABO1 and ABO2, respectively.
In the similar manner, the parameter set denoted as BLW \cite{Braun:2006hz} has been determined.

The Borel parameter $M^2$ corresponds to the inverse imaginary time (squared)
at which matching of the QCD calculation is done to the expansion in hadronic states.
Usually, we can try to take $M^2$ as small as possible in order to reduce sensitivity to the contributions
of higher-mass states, which is the main irreducible uncertainty of the LCSRs method.
Thus, despite our analysis is limited by the leading order, in this work as in Ref.~\cite{Anikin:2013aka}, we take $M^2=1.5$~GeV$^2$ and  $M^2=2$~GeV$^2$ as two acceptable choices.

\begin{figure}[ht]
\vspace{0.3cm}
\includegraphics[width=0.35\textwidth]{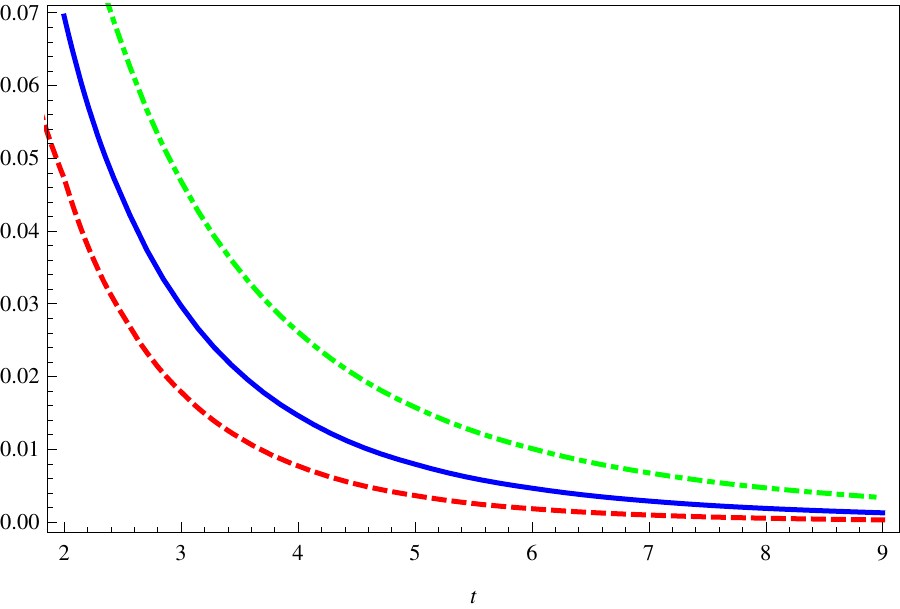}
  \caption{The sum of gravitational form factors $\mathds{A}(t)$  and $\mathds{B}(t)$ which gives the first moment of $H(x, \xi, t) + E(x, \xi, t)$
for the region of large $t$.
  Notations: the solid blue line corresponds to ABO2; the dashed red line corresponds to ABO1; the dot-dashed green
  line corresponds to BLW \cite{Anikin:2013aka}.}
\label{Fig-3}
\end{figure}

As well-known, the gravitational form factors $\mathds{A}(t)$ and $\mathds{B}(t)$ define the Mellin moment
of generalized parton distributions as
\begin{eqnarray}
\label{sumA-B}
\int\limits_{-1}^{1} dx\, x \, \Big( H(x, \xi, t) + E(x, \xi, t)\Big) = \mathds{A}(t) + \mathds{B}(t)=2\mathds{J}(t)
\end{eqnarray}
which directly relates to the total nucleon spin.
Our results for the sum of $\mathds{A}(t)$ and $\mathds{B}(t)$ with the valence quark combinations are depicted in Fig.~\ref{Fig-3}.

The normalization condition given by
\begin{eqnarray}
\label{Norm-A}
\sum_{a=q,g,...} \mathds{A}^a(0)=1
\end{eqnarray}
takes place provided one sums over all parton contributions inside the nucleon including the valence, sea and other contributions
to form the nucleon spin \footnote{In Ref.~\cite{Obukhov:2013zca}, the particle spin has been considered in an arbitrary gravitation field.}.
Our LCSR calculations show that the fitting parameter $\kappa$ of (\ref{A-fit}) is close to $1$.
It means that for more comprehensive analysis it is necessary to take into account the next-to-leading-order corrections to the LCSRs
together with the gluon and sea quark contributions. We plan to study this in the future works.

%
%
%

\subsection{Plus-minus light-cone projections of the amplitude: estimated form factor $ \mathds{D}(t)$}
\label{Subsec:plus-minus-proj}

We are now going over to the plus-minus light-cone projection of the amplitude which is $T^{+-}(P,\Delta)$.
Introducing the combination and the notations given by
\begin{eqnarray}
\label{F}
&&\mathds{F}(t)= 2m_N\Big(m_N\overline{\mathds{C}}(t) - \frac{\Delta^2_\perp}{4m_N}\mathds{D}(t) \Big),
\nonumber\\
&&T^{+-}_{1(4)\,[\bar\psi_\pm \psi_\pm]}=
\bar T^{+-}_{1(4)\,[\bar\psi_\pm \psi_\pm]} N^+(P),
\end{eqnarray}
the first type of amplitudes yields
\begin{eqnarray}
\label{Pol-sr-1}
&&\hspace{-0.6cm}\frac{\lambda_1 m_N}{m^2_N - P^{\prime\,2}}
\Big(
m^2_N \mathds{A}(t) + \Big[ \frac{\Delta^2_\perp}{4} + P\cdot\Delta \Big]\mathds{A}(t)
+ \frac{\Delta^2_\perp}{4}\mathds{B}(t)
\nonumber\\
&&\hspace{-0.6cm}
 + \mathds{F}(t)
\Big) N^+(P)
=
m_N \Big(
\bar T^{+-}_{1\,[\bar\psi_+ \psi_+]} + \bar T^{+-}_{1\,[\bar\psi_- \psi_-]}
\Big) N^+(P)
\end{eqnarray}
where
\begin{eqnarray}
\label{Pol-sr-1-2}
&&\hspace{-0.5cm}m_N \Big(
\bar T^{+-}_{1\,[\bar\psi_+ \psi_+]} + \bar T^{+-}_{1\,[\bar\psi_- \psi_-]}
\Big)=m_N \bar T^{+-}_{1\,[\bar\psi_- \psi_-]} +
\\
&&\hspace{-0.5cm}
\frac{m_N }{\pi}\int_{0}^{s_0} \frac{ds}{s-P^{\prime\,2}} \text{Im}\Big\{
\frac{m^2_N}{4} \mathcal{A}^{++}(s,t) + \frac{P\cdot\Delta}{4}\mathcal{A}^+_{em}(s,t)
\Big\}.
\nonumber
\end{eqnarray}
For the second type of amplitudes, we can write the following
\begin{eqnarray}
\label{Pol-sr-2}
&&\frac{\lambda_1}{m^2_N - P^{\prime\,2}}
\Big(
\frac{m^2_N}{2} \mathds{A}(t) - \frac{P\cdot\Delta}{2}\mathds{B}(t) + \frac{1}{2} \mathds{F}(t)
\Big) \hat\Delta_\perp N^+(P)
\nonumber\\
&&
=
\hat\Delta_\perp \Big(
\bar T^{+-}_{4\,[\bar\psi_+ \psi_+]} + \bar T^{+-}_{4\,[\bar\psi_- \psi_-]}
\Big)  N^+(P)
\end{eqnarray}
where
\begin{eqnarray}
\label{Pol-sr-2-2}
&&\hspace{-0.5cm}\,\hat\Delta_\perp \Big(
\bar T^{+-}_{4\,[\bar\psi_+ \psi_+]} + \bar T^{+-}_{4\,[\bar\psi_- \psi_-]}
\Big)=\hat\Delta_\perp\, \bar T^{+-}_{4\,[\bar\psi_- \psi_-]} +
\\
&&\hspace{-0.5cm}
\frac{\hat\Delta_\perp}{\pi}\int_{0}^{s_0} \frac{ds}{s-P^{\prime\,2}} \text{Im}\Big\{
\frac{m^2_N}{4} \mathcal{B}^{++}(s,t) + \frac{P\cdot\Delta}{4}\mathcal{B}^+_{em}(s,t)
\Big\} .
\nonumber
\end{eqnarray}
Unlike the plus-plus light-cone projection, the eqns.~(\ref{Pol-sr-1}) and (\ref{Pol-sr-2}) do not form
the closed system of equations. Hence, there is no way to compute the form factors $\mathds{D}(t)$
directly. Instead, we shall try to estimate the form factor $\mathds{D}(t)$.

The next step is to perform the Borel transforms of eqns.~(\ref{Pol-sr-1}) and (\ref{Pol-sr-2}).
Afterwards using eqn.~(\ref{A-B-ff}),
we derive the first sum rules which read
\begin{eqnarray}
\label{pm-proj-1}
&&\mathds{F}(t)=\bar T^{+-}_{1\,[\bar\psi_- \psi_-]} -
\int \hat d\mu(s)\Big\{ \frac{m^2_N}{4} \mathcal{A}^{++} -
\nonumber\\
&&
 \frac{\Delta^2_\perp}{4} \Big(
\frac{1}{2} \mathcal{A}^{++} + \mathcal{B}^{++} - \frac{1}{2} \mathcal{A}^+_{em}
\Big)
\Big\}
\end{eqnarray}
and the second sum rules which are
\begin{eqnarray}
\label{pm-proj-2}
&&\mathds{F}(t)=2\,\bar T^{+-}_{4\,[\bar\psi_- \psi_-]}
- \int \hat d\mu(s)\Big\{ \frac{m^2_N}{2}\Big( \mathcal{A}^{++} - \mathcal{B}^{++}\Big)
-
\nonumber\\
&&
\frac{\Delta^2_\perp}{4} \Big(
2\mathcal{B}^{++} - \mathcal{B}^+_{em}
\Big) \Big\}.
\end{eqnarray}
Eqns.~(\ref{pm-proj-1}) and (\ref{pm-proj-2}) yield the following integral relation:
\begin{eqnarray}
\label{pm-proj-f}
&&
\Big( 1+ \frac{2m^2_N}{\Delta^2_\perp} \Big)\, \int \hat d\mu(s) \Big\{ \frac{\mathcal{A}^{++}}{2}  - \mathcal{B}^{++} \Big\}
=
\nonumber\\
&&
 \int \hat d\mu(s) \Big\{ \frac{\mathcal{A}^{+}_{em} }{2}- \mathcal{B}^+_{em}\Big\}
+ \frac{4}{\Delta^2_\perp}\bar T^{+-}_{41\,[\bar\psi_- \psi_-]}
\end{eqnarray}
where, by definition,
\begin{eqnarray}
 \bar T^{+-}_{41\,[\bar\psi_- \psi_-]} =  2 \bar T^{+-}_{4\,[\bar\psi_- \psi_-]} -  \bar T^{+-}_{1\,[\bar\psi_- \psi_-]}.
\end{eqnarray}
Based on the relation (\ref{pm-proj-f}), it is instructive to analyse the functions $R_1(t)$ and $R_2(t)$ given by
\begin{eqnarray}
\label{R1}
R_1(t)=\frac{\int \hat d\mu(s) \Big\{ \frac{1}{2} \mathcal{A}^{++} - \mathcal{B}^{++}\Big\}}
{\int \hat d\mu(s) \Big\{ \frac{1}{2}\mathcal{A}^{+}_{em} - \mathcal{B}^+_{em}\Big\}}
\Big( 1+ \frac{2m^2_N}{\Delta^2_\perp} \Big) - 1
\end{eqnarray}
and
\begin{eqnarray}
\label{R2}
&&R_2(t)=
\int \hat d\mu(s) \Big\{
\frac{\mathcal{A}^{++} }{2} - \mathcal{B}^{++} \Big\}\Big( 1+ \frac{2m^2_N}{\Delta^2_\perp} \Big)
\nonumber\\
 &&-
\int \hat d\mu(s) \Big\{ \frac{\mathcal{A}^{+}_{em}}{2} - \mathcal{B}^+_{em}\Big\}.
\end{eqnarray}
%
\begin{figure}[ht]
\vspace{0.3cm}
\includegraphics[width=0.35\textwidth]{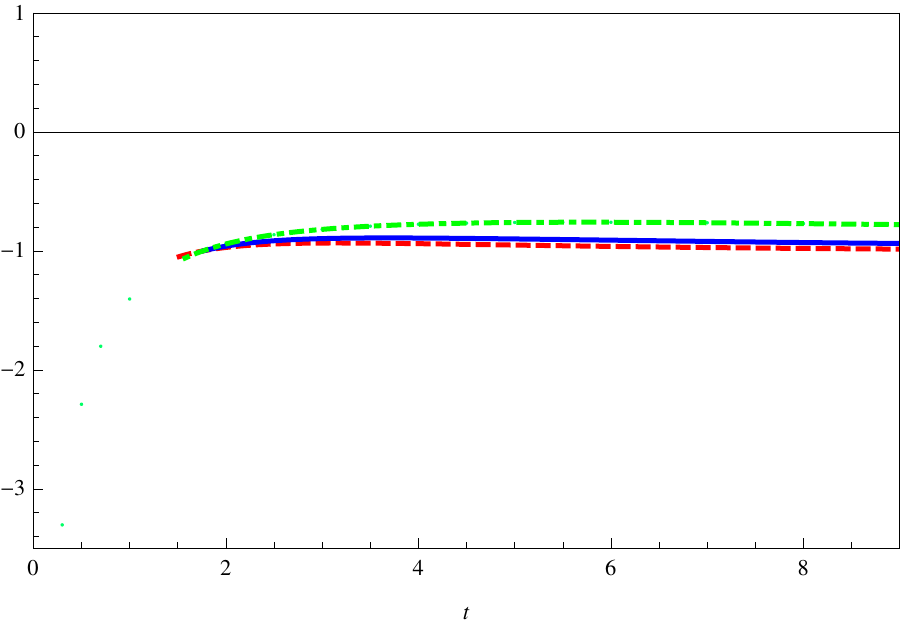}
\\
\vspace{0.3cm}
\includegraphics[width=0.35\textwidth]{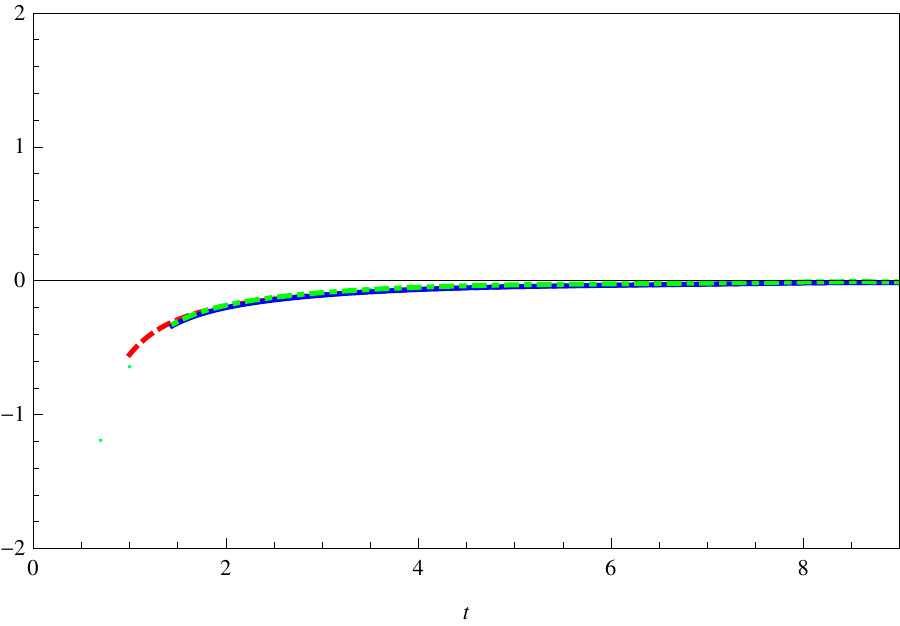}
  \caption{The functions $R_1(t)$ (the upper panel)
  and $R_2(t)$ (the lower panel).
  Notations: the solid blue line corresponds to ABO2; the dashed red line corresponds to ABO1; the dot-dashed green
  line corresponds to BLW \cite{Anikin:2013aka}.}
\label{Fig-R}
\end{figure}
%
The behaviour of $R_1$ and $R_2$ as functions of $t$ has been presented in Fig.~\ref{Fig-R}.
The numerical analysis shows that $R_1$ and $R_2$ scale approximately $-0.8$ and $-0.02$, respectively.
If we reckon that  $R_2\approx 0$ and $R_1\approx -1$, we are able to estimate the unknown LCSRs contributions as
\begin{eqnarray}
\label{Est}
 2 \lim_{t\to\infty}\bar T^{+-}_{4\,[\bar\psi_- \psi_-]} \approx \lim_{t\to\infty} \bar T^{+-}_{1\,[\bar\psi_- \psi_-]}
\end{eqnarray}
and
\begin{eqnarray}
\label{Est-2}
&& \lim_{t\to\infty} \frac{4}{\Delta^2_\perp}\Big( 2\,\bar T^{+-}_{4\,[\bar\psi_- \psi_-]} - \bar T^{+-}_{1\,[\bar\psi_- \psi_-]} \Big) \approx
\nonumber\\
&&
\lim_{t\to\infty}\int \hat d\mu(s) \Big\{ \mathcal{B}^+_{em} -  \frac{1}{2}\mathcal{A}^{+}_{em} \Big\}.
\end{eqnarray}
Our numerical analysis shows that these estimations work with a good precision for $t\geq 2\, \text{GeV}^2$.

Since the amplitudes of $\mathcal{A}$- and $\mathcal{B}$-types are nothing but the invariant amplitudes of  $T_1$- and  $T_4$-types
(see eqns.~(\ref{Pol-sr-1-2}) and (\ref{Pol-sr-2-2})), from eqns.~ (\ref{pm-proj-f}), (\ref{Est}) and (\ref{Est-2}) 
we group together the $\mathcal{A}$ ($\mathcal{B}$)-type of form factors for $T_1$ ($T_4$)-type of amplitude and, therefore,
we impose that
 \begin{eqnarray}
\label{Est-3}
&&\lim_{t\to\infty}\, \frac{4}{\Delta^2_\perp}\, \bar T^{+-}_{4\,[\bar\psi_- \psi_-]} \approx
\\
&&
\lim_{t\to\infty} \int \hat d\mu(s) \Big\{ \frac{1}{2}\mathcal{B}^+_{em} -
\frac{1}{2}\mathcal{B}^{++}\Big( 1- \frac{2m^2_N}{t} \Big)  \Big\}\approx
\nonumber\\
&&\lim_{t\to\infty} \int \hat d\mu(s) \Big\{ \frac{1}{2}\mathcal{B}^+_{em}\Big\}
\nonumber
\end{eqnarray}
and
\begin{eqnarray}
\label{Est-3-2}
&&\lim_{t\to\infty} \, \frac{4}{\Delta^2_\perp}\, \bar T^{+-}_{1\,[\bar\psi_- \psi_-]} \approx
\\
&&
\lim_{t\to\infty} \int \hat d\mu(s) \Big\{ \frac{1}{2}\mathcal{A}^+_{em} -
\frac{1}{2}\mathcal{A}^{++}\Big( 1- \frac{2m^2_N}{t} \Big)  \Big\}\approx
\nonumber\\
&&\lim_{t\to\infty} \int \hat d\mu(s) \Big\{ \frac{1}{2}\mathcal{A}^+_{em}\Big\}
\nonumber
\end{eqnarray}
provided $|\mathcal{B}^+_{em}|\gg |\mathcal{B}^{++}\Big( 1- 2m^2_N/t \Big)|$ and
$|\mathcal{A}^+_{em}|\gg |\mathcal{A}^{++}\Big( 1- 2m^2_N/t \Big)|$ 
which are not at odds with the
numerical analysis. 
Indeed, even for the region of moderate $t$, {\it i.e.} $t\in [7,\, 10] \, \text{GeV}^2$, we have 
$|\mathcal{B}^+_{em}| / |\mathcal{B}^{++}\Big( 1- 2m^2_N/t \Big)|\sim 0.3 $ while 
$|\mathcal{A}^+_{em}| / |\mathcal{A}^{++}\Big( 1- 2m^2_N/t \Big)| \sim 0.1$.

\begin{figure}[ht]
\vspace{0.3cm}
\includegraphics[width=0.35\textwidth]{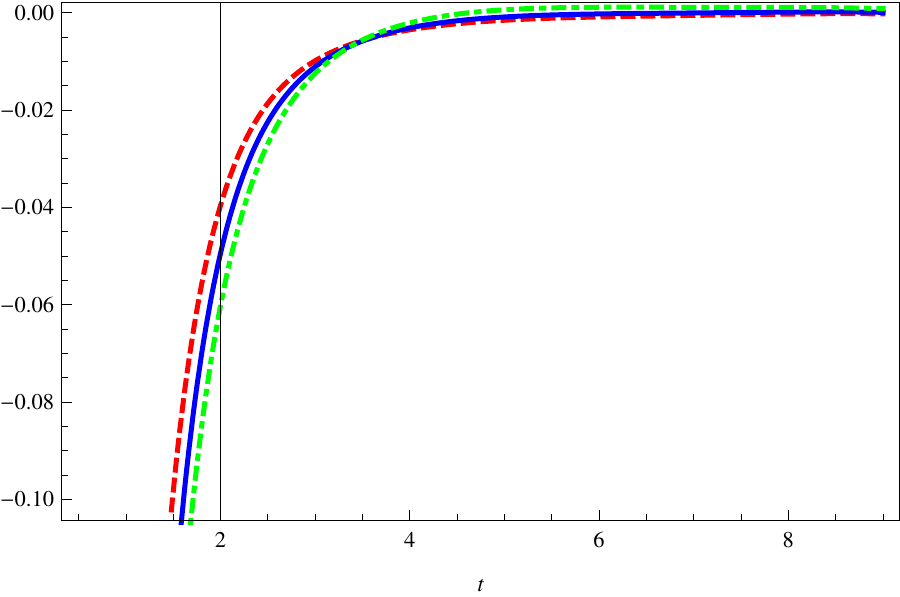}
\\
\vspace{0.3cm}
\includegraphics[width=0.35\textwidth]{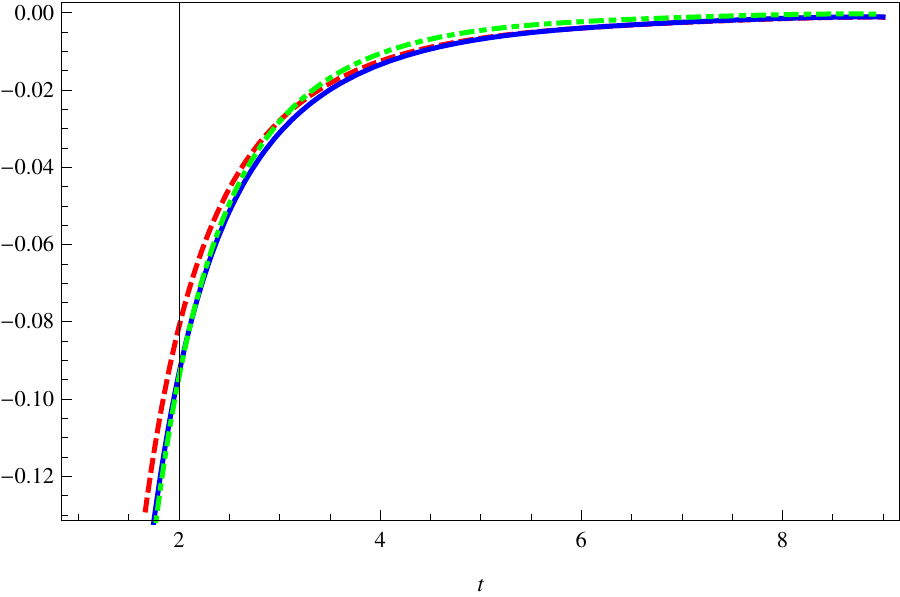}
  \caption{The  gravitational form factor $\mathds{D}(t)$ for the region of large $t$:
$\frac{5}{4}\mathds{D}_{4\, [\bar\psi_+ \psi_+]}(t)$ (the upper panel)
  and the estimated total $\frac{5}{4}\mathds{D}(t)$ (the lower panel).
  Notations: the solid blue line corresponds to ABO2; the dashed red line corresponds to ABO1; the dot-dashed green
  line corresponds to BLW \cite{Anikin:2013aka}.}
\label{Fig-D}
\end{figure}

We now analyse the $D$-term contributions. Neglecting $\overline{\mathds{C}}$-term from
$\mathds{F}$-combination at the large $t$,
we first derive the $[\bar\psi_+ \psi_+]$-contribution to the $D$-term form factor.
For example, from eqn.~(\ref{pm-proj-2}), we have
\begin{eqnarray}
\label{D-FF-1}
&&-2 \mathds{D}_{4\, [\bar\psi_+ \psi_+]} \approx
\\
&&
\int \hat d\mu(s) \Big\{
2\mathcal{B}^{++} - \mathcal{B}^+_{em}
-\frac{2m^2_N}{\Delta^2_\perp}\Big(\mathcal{A}^{++} - \mathcal{B}^{++}\Big) \Big\}.
\nonumber
\end{eqnarray}
On the other hand, inserting eqn.~(\ref{Est-3}) into eqn.~(\ref{pm-proj-2}) or eqn.~(\ref{Est-3-2}) into eqn.~(\ref{pm-proj-1}),
we estimate the
form factor $\mathds{D}(t)$ by
\begin{eqnarray}
\label{Dfull-Est}
\mathds{D}(t) \approx
 - \int \hat d\mu(s) \Big\{
\frac{1}{2}\mathcal{B}^{++} +
\frac{m^2_N}{t} \mathcal{A}^{++} \Big\}.
\end{eqnarray}
We emphasize that the form factor $\mathds{D}(t)$ is independent one.
Eqn.~(\ref{Dfull-Est}) merely tell us that the estimated $\mathds{D}(t)$ as a function of $t$ behaves like
a combination of the form factors $\mathds{A}(t)$ and  $\mathds{B}(t)$.
The results of the calculations for the estimated $\frac{5}{4}\mathds{D}$
and $\frac{5}{4}\mathds{D}_{4\, [\bar\psi_+ \psi_+]}$ have been demonstrated in Fig.~\ref{Fig-D}.
It is important to note that the sign of $D$-term form factor is mostly determined by the sign of $\mathds{A}(t)$
rather than by sign of $\mathds{B}(t)$ in the region where $t$ is relatively moderate one.

\begin{figure}[ht]
\vspace{0.3cm}
\includegraphics[width=0.35\textwidth]{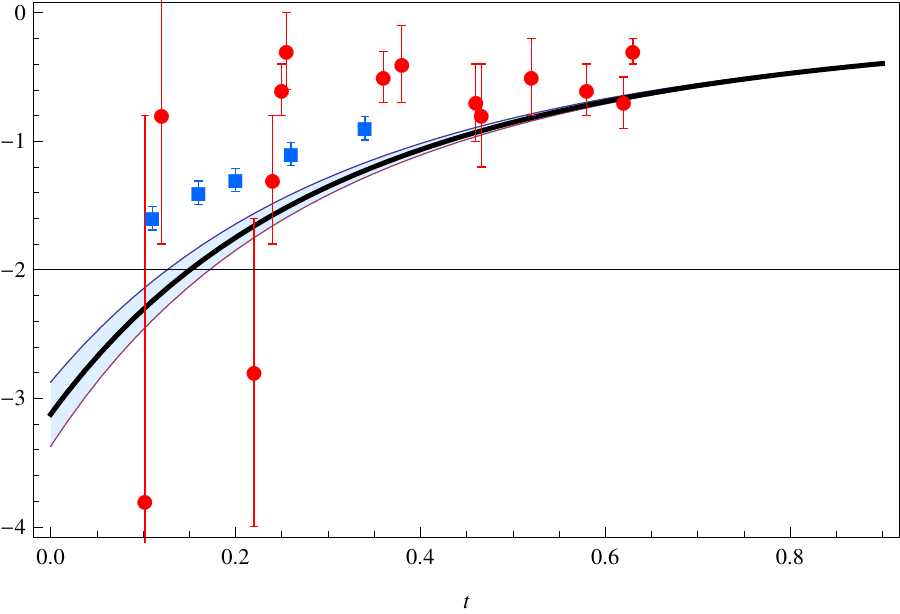}
  \caption{The gravitational form factor $\frac{5}{4}\mathds{D}^{\text{fit}}(t)$ which is analytically continued to the region of small $t$.
The experimental data: the blue squares -- the Jefferson Lab \cite{Burkert:2018bqq}, the red bullet points --
the Lattice QCD \cite{Hagler:2007xi}.}
\label{Fig-D-fit}
\end{figure}

In the same manner as in the case of the form factors $\mathds{A}$ and $\mathds{B}$,
the estimated $\mathds{D}(t)$ can be approximated by (cf. \cite{Tanaka:2018wea})
\begin{eqnarray}
\label{Dtot-fit}
&&\mathds{D}^{\text{fit}}(t) = \frac{D}{(1+e\, t)^n},
\nonumber\\
&&
D=-2.5\pm 0.2, \quad e=0.93\pm 0.03\,\text{GeV}^{-2},
\nonumber\\
&&n= 3.4\pm 0.025.
\end{eqnarray}
In the small $t$-region, for all sets of LCSR-parametrizations, this approximating function matches, see Fig.~\ref{Fig-D-fit}, with the first experimental data
\cite{Burkert:2018bqq,Hagler:2007xi}
together with the results of the chiral quark-soliton and Skyrme models presented in \cite{Polyakov:2018zvc}.
Our function $\mathds{D}^{\text{fit}}(t)$ lies below than the quark-soliton prediction and above than the result of Skyrme
model.
%
%
%

\subsection{Mechanical properties of nucleon}
\label{Subsec:Mech}

Having calculated the form factors $\mathds{D}(t)$ and $\mathds{A}(t)$, one computes the pressure, the energy density in the
center of nucleon and the hadron mechanical radius \cite{Polyakov:2018zvc}. We have
\begin{eqnarray}
\label{Pres}
p_0\hspace{-0.3cm}&&=-\frac{1}{24\pi^2 m_N} \int^{\infty}_{0} dt t \sqrt{t} \mathds{D}^{\text{fit}}(t)
\nonumber\\
&&
= 0.84 \,\text{GeV}/\text{fm}^{3} \,\,\, [\text{cf.}\,\, p_0^{\text{\cite{Polyakov:2018zvc}}}=0.23 \,\text{GeV}/\text{fm}^{3}]
\end{eqnarray}
for the pressure, and
\begin{eqnarray}
\label{Eden}
{\cal E}\hspace{-0.3cm}&&=\frac{m_N}{4\pi^2} \int^{\infty}_{0} dt \sqrt{t}
\Big( \mathds{A}^{\text{fit}}(t) +
\frac{t}{4m^2_N} \mathds{D}^{\text{fit}}(t) \Big)
\nonumber\\
&&
= 0.92 \,\text{GeV}/\text{fm}^{3} \,\,\, [\text{cf.}\,\, {\cal E}^{\text{\cite{Polyakov:2018zvc}}}=1.7 \,\text{GeV}/\text{fm}^{3}]
\end{eqnarray}
for the energy density, and
\begin{eqnarray}
\label{Rmech}
\langle r^2\rangle_{\text{mech}}=6D \Big[ \int^{\infty}_{0} dt \mathds{D}^{\text{fit}}(t)\Big]^{-1}
= 13.4 \,\text{GeV}^{-2}
\end{eqnarray}
for the mechanical radius.

\begin{figure}[ht]
\vspace{0.3cm}
\includegraphics[width=0.35\textwidth]{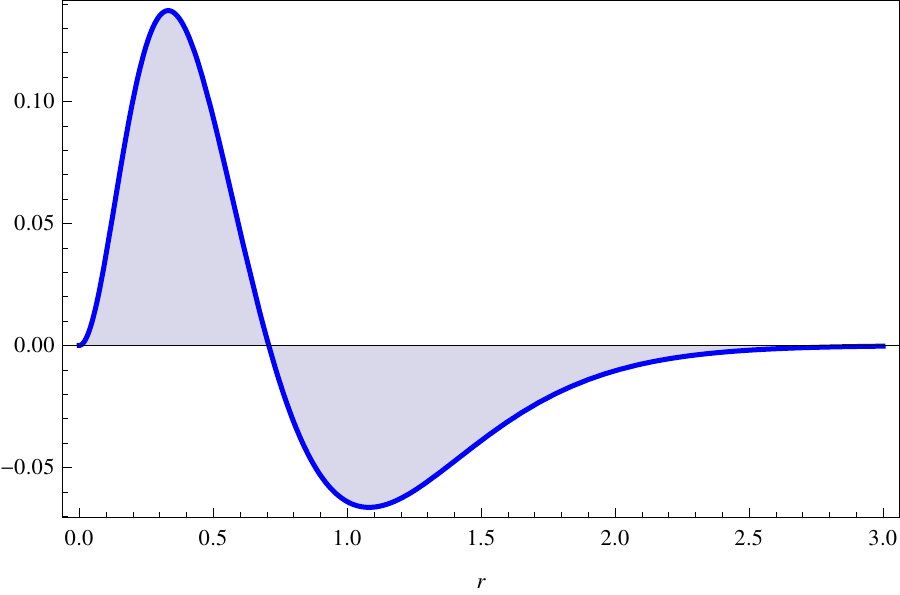}
  \caption{The normalized pressure $4 \pi\, m_N \, r^2 p(r)\, [\text{GeV/fm}]$.}
\label{Pre-Fun-Fig}
\end{figure}

\begin{figure}[ht]
\vspace{0.3cm}
\includegraphics[width=0.35\textwidth]{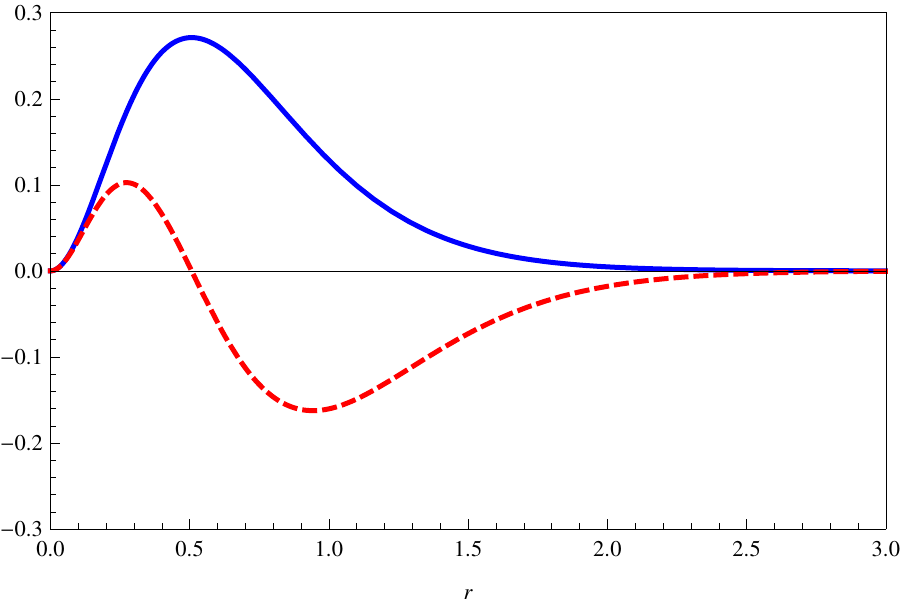}
  \caption{The normal forces $F_n$ (the solid line) and tangential forces
$F_t$ (the dashed line) in $[\text{GeV/fm}]$.}
\label{Fn-Ft-Fig}
\end{figure}

Moreover, following \cite{Polyakov:2018zvc},
we present the result, see Fig.~\ref{Pre-Fun-Fig}, for the pressure $p(r)$ as a function of $r=|\vec{\bf r}_\perp|$ given by
\begin{eqnarray}
\label{Pre-fun}
&&p(r)=\frac{1}{6m_N}\frac{1}{r^2}\frac{d}{dr} r^2 \frac{d}{dr} \widetilde{\mathds{D}}^{\text{fit}}(r),
\nonumber\\
&& \widetilde{\mathds{D}}^{\text{fit}}(r)=\frac{1}{(2\pi)^3}\int d^2\vec{\Delta}_\perp
e^{-i \vec{\bf r}_\perp \vec{\Delta}_\perp } \mathds{D}^{\text{fit}}(\vec{\Delta}_\perp^2).
\end{eqnarray}
One can see that the normalized pressure $4\pi\, m_N\, r^2\, p(r)$ has zero at $r=0.7 \,\text{fm}$.
The computed function $p(r)$ meets the von Laue condition, {\it i.e.}
\begin{eqnarray}
\label{Laue}
\int_{0}^{\infty} dr\, r^2\, p(r) = 0,
\end{eqnarray}
which is a consequence of the EMT conservation and it shows how the internal forces balance inside a composed particle
\cite{Polyakov:2002yz, Polyakov:2018zvc}.
In addition, the spherical shell of radius $r$ in the nucleon are undergoing the normal and tangential
forces: $F_n$ and $F_t$, respectively.
With the obtained $\widetilde{\mathds{D}}^{\text{fit}}(r)$, we can readily calculate these forces:
\begin{eqnarray}
\label{Fn-Ft}
&&F_n=4\pi m_N r^2 \Big( p(r) + \frac{2}{3} s(r)\Big),
\nonumber\\
&&F_t=4\pi m_N r^2 \Big( p(r) - \frac{1}{3} s(r)\Big)
\end{eqnarray}
where the shear forces $s(r)$ reads
\begin{eqnarray}
\label{Sh}
s(r)=-\frac{1}{4m_N} r \frac{d}{dr} \frac{1}{r} \frac{d}{dr} \widetilde{\mathds{D}}^{\text{fit}}(r).
\end{eqnarray}
Fig.~\ref{Fn-Ft-Fig} shows us that the estimated normal and tangential forces which correspond to the
valence quark combination are considerably small. This is fairly concordant with the chiral quark-soliton model
as well.
%
%
%

\subsection{Plus-perp light-cone projections of the amplitude}
\label{Subsec:plus-perp-proj}

The last stage is the consideration of the plus-perp light-cone projection of the amplitude, $T^{+\perp}(P,\Delta)$,  it gives us
\begin{eqnarray}
\label{Pol-sr-3}
\frac{\lambda_1 m_N \Delta^2_\perp}{m^2_N - P^{\prime\,2}}
 \mathds{A}(t) N^+(P) =
m_N\Big(
\bar T^{+\perp}_{1\,[\bar\psi_+ \psi_+]} + \bar T^{+\perp}_{1\,[\bar\psi_+ \psi_-]}
\Big) N^+(P)
\nonumber\\
\end{eqnarray}
where
\begin{eqnarray}
\label{Pol-sr-3-2}
&&\hspace{-0.5cm}m_N \Big(
\bar T^{+\perp}_{1\,[\bar\psi_+ \psi_+]} + \bar T^{+\perp}_{1\,[\bar\psi_+ \psi_-]}
\Big)=m_N \bar T^{+\perp}_{1\,[\bar\psi_+ \psi_-]} +
\nonumber\\
&&\hspace{-0.5cm}
\frac{m_N }{\pi}\int_{0}^{s_0} \frac{ds}{s-P^{\prime\,2}} \text{Im}\Big\{
\frac{\Delta^2_\perp}{4}\mathcal{A}^+_{em}\Big\}
\end{eqnarray}
and
\begin{eqnarray}
\label{Pol-sr-4}
&&\frac{\lambda_1}{m^2_N - P^{\prime\,2}}
\Big( P\cdot\Delta \mathds{J}(t) +
\frac{\Delta^2_\perp}{4} \big[\mathds{A}(t)-\mathds{B}(t)\big]
\Big) \hat\Delta_\perp N^+(P)
\nonumber\\
&&
=\hat\Delta_\perp\Big(
\bar T^{+\perp}_{4\,[\bar\psi_+ \psi_+]} + \bar T^{+\perp}_{4\,[\bar\psi_+ \psi_-]}
\Big) N^+(P)
\end{eqnarray}
where
\begin{eqnarray}
\label{Pol-sr-4-2}
&&\hspace{-0.5cm}\hat\Delta_\perp \Big(
\bar T^{+\perp}_{4\,[\bar\psi_+ \psi_+]} + \bar T^{+\perp}_{4\,[\bar\psi_+ \psi_-]}
\Big)=\hat\Delta_\perp \bar T^{+\perp}_{4\,[\bar\psi_+ \psi_-]} +
\nonumber\\
&&\hspace{-0.5cm}
\frac{\hat\Delta_\perp }{\pi}\int_{0}^{s_0} \frac{ds}{s-P^{\prime\,2}} \text{Im}\Big\{
\frac{\Delta^2_\perp}{4}\mathcal{B}^+_{em}\Big\}
\end{eqnarray}
After the Borel transforms and using eqn.~(\ref{A-B-ff}), we obtain
\begin{eqnarray}
\label{F-1}
&&\hspace{-0.8cm}\int \hat d\mu(s) \Big\{ \frac{1}{2}\mathcal{A}^{++} \Big\}
= \int \hat d\mu(s) \Big\{ \frac{1}{4}\mathcal{A}^{+}_{em} \Big\}+ 
\frac{4}{\Delta^2_\perp}\bar T^{+\perp}_{1\,[\bar\psi_+ \psi_-]},
\\
&&\hspace{-0.8cm}\int \hat d\mu(s) \Big\{ \frac{1}{2}\mathcal{B}^{++} \Big\} =
\int \hat d\mu(s) \Big\{ \frac{1}{4}\mathcal{B}^{+}_{em} \Big\} + 
\frac{4}{\Delta^2_\perp}\bar T^{+\perp}_{4\,[\bar\psi_+ \psi_-]}.
\end{eqnarray}
These sum rules allow us to calculate the contributions of $\bar T^{+\perp}_{1, 4\, [\bar\psi_+ \psi_-]}$ explicitly.

\section{Conclusions and discussions}
\label{Sec:Conclusions}

The main challenge in the QCD description of any form factors is the calculation of soft contributions which
correspond to the Feynman mechanism to transfer the rather large momentum.
As demonstrated in many papers (see, for example, \cite{Balitsky:1986st, Balitsky:1989ry, Braun:2006hz, Braun:2001tj})
the LCSRs approach is attractive due to the fact that the soft contributions are calculated in terms of the same
DAs that the pQCD calculations include.
Thus, the LCSRs can be positioned as one of the most direct relations of the different kind of form
factors and hadron DAs that is available at present.

In the paper, we have presented the first calculations and several estimations for
the gravitational form factors implemented due to the LCSRs techniques at the leading order.

We have shown that the essential contributions to the gravitational form factors can be computed owing
to the suitable adoption of the LCSRs designed for the electromagnetic form factors.
Within the developed approach, focusing on the valence quark content of nucleon,
we have directly computed the gravitational form factors $\mathds{A}(t)$  and $\mathds{B}(t)$
for sufficiently large Euclidian $t\gtrsim 1\, \text{GeV}^2$ where the LCSRs approach is reliable.
For this region, we have approximated the obtained form factors by the multipole functions
$\mathds{A}^{\text{fit}}(t)$  and $\mathds{B}^{\text{fit}}(t)$ and, then, have analytically continued to
the small $t$ region.  With the calculated form factors, we have presented the result for
the Mellin $x$-moment of GPDs combination $H(x, \xi, t) + E(x, \xi, t)$.

Also, the several estimations and predictions for the valence quark contributions to
the gravitational form factor $\mathds{D}(t)$ have been done.
To the large $t$ region where the LCSRs approach can be used,
we have found that the estimated form factor $\mathds{D}$ as a function of $t$ has
the similar behaviour as the suitable combination of the form factors $\mathds{A}(t)$ and $\mathds{B}(t)$.
Regarding the small $t$ region, where the first experimental data are available \cite{Hagler:2007xi, Burkert:2018bqq},
the estimated $\mathds{D}(t)$ can be approximated by the fitting multipole function
$\mathds{D}^{\text{fit}}(t)$. With the estimated $\mathds{D}^{\text{fit}}(t)$, a few
quantities characterizing the so-called mechanical properties of the nucleon
have been calculated. We emphasize that all
our results are {\it not} at odds with the experimental data
as well as with the results of the chiral quark-soliton and Skyrme models presented in \cite{Polyakov:2018zvc}.

\section*{Acknowledgements}
We are grateful to V.~Braun, M.~Deka, A.~Manashov, M.~Polyakov, P.~Schweitzer and O.~Teryaev for useful discussions,
comments and remarks.
The special thanks go to M.~Vanderhaeghen for the attraction of our attention to
the gravitational form factors.
This work was supported in part by
the Heisenberg-Landau Program.

\appendix
\renewcommand{\theequation}{\Alph{section}.\arabic{equation}}

\section{Collinear and geometrical twists}
\label{App}

For pedagogical reasons, we recall briefly the main items from the theory of collinear and geometrical
twists.
We begin with the non-local gauge-invariant quark operator which can be expanded as
\begin{eqnarray}
\label{non-loc}
\bar\psi(0)\gamma^\mu \,[0\,;\,z]_A\, \psi(z)=
\sum\limits_{n=0}^{\infty} \frac{1}{n!} z_{\alpha_1}...z_{\alpha_n} {\cal O}^{\mu\alpha_1...\alpha_n}(0),
\end{eqnarray}
where ($\vec{\cal D}^{\alpha}=\vec{\partial}^\alpha-igA^\alpha$)
\begin{eqnarray}
\label{wl-1}
[0\,;\,z]_A = \mathbb{P}\text{exp}\Big\{ig\int\limits_{z}^{0} d\omega_{\nu}A^\nu(\omega) \Big\}
\end{eqnarray}
and
\begin{eqnarray}
\label{Oper-1}
{\cal O}^{\mu\alpha_1...\alpha_n}(0)=\bar\psi(0)\gamma^\mu \vec{\cal D}^{\alpha_1}...\vec{\cal D}^{\alpha_n} \psi(0).
\end{eqnarray}
Here, all of the Lorentz indices are free and ${\cal O}^{\mu\alpha_1...\alpha_n}$ does not belong to the irreducible representation ${SO(3,1)}$.
The Lorentz rank-$(n+1)$ tensor can be decomposed over the irreducible representations with a help of the Young tableaux
(see, for example, \cite{Belitsky:2005qn}).

To describe the hard inclusive and exclusive processes, it is instructive to distinguish the geometrical and collinear twists:
\begin{eqnarray}
\label{def-tws}
&&\text{geom. twist $\tau$} = \text{dimension $d$} - \text{spin $s$},
\\
&&\text{coll. twist $t$} = \text{dimension $d$} - \text{spin projection $s_a$}.
\nonumber
\end{eqnarray}
Notice that the definite Lorentz spin $s$ can be associated with the local operators only,
while the certain Lorentz spin projection $s_a$ can be related to for both the local and non-local operators.
Indeed, the certain spin projection, say, $s_a=+1$ can be corresponded to many values of spin: $s=+1, +2, +3$ etc.

Despite the fact that the irreducible representations of the Lorentz group and its collinear subgroup
can only be realized with the local quark-gluon operators, in many cases (for example, in the exclusive process case) it is convenient to introduce the collinear twist for the subjects which are not forming the irreducible representations. For instance, the individual light-cone spinors
$\psi_{\pm}(0)$ have the collinear twists $t=1$ for the plus projection and $t=2$ for the minus projection, and
$\bar\psi(0)\gamma^\pm \psi(z)$  and $\bar\psi(0)\gamma^\perp \psi(z)$ correspond to the collinear twists $t=2(4)$ and $t=3$, respectively, and so on.
We remind that, on the basis of the Weyl representation, the (anti)quarks belong to the spinorial representations, $(\frac{1}{2},0)$ or $(0,\frac{1}{2})$,
while the Lorentz vectors form the vector representations, $(\frac{1}{2},\frac{1}{2})$.

Thus, the geometrical twist-$2$ operators is defined as
\begin{eqnarray}
\label{Op-tw2}
&&{\cal R}^{\mu\alpha_1...\alpha_n}_{\tau=2}= \text{\Large\bf S}_{(\text{\footnotesize all})}\,
\bar\psi(0)\gamma^\mu \vec{\cal D}^{\alpha_1}...\vec{\cal D}^{\alpha_n} \psi(0)
\nonumber\\
&&
\equiv{\cal O}^{\{\mu\alpha_1...\alpha_n\}}(0) - \text{trace},
\end{eqnarray}
where we introduce the symmetrization symbol which acts, for instance, on the rank-$2$ local operator as
\begin{eqnarray}
\text{\Large\bf S}_{(\mu\nu)}\, {\cal O}^{\mu\nu}  =
\frac{1}{2!} \left( {\cal O}^{\mu\nu} + {\cal O}^{\nu\mu} - \frac{1}{2} g^{\mu\nu} g_{\alpha\beta} {\cal O}^{\alpha\beta} \right).
\end{eqnarray}
It is easy to show that the symmetrization procedure can be presented in the following form:
\begin{eqnarray}
\label{Dif-sym}
{\cal O}^{\{\mu\alpha\}}(0) =\frac{1}{2} \lim_{z\to 0} \frac{\partial^2}{\partial z^\mu \partial z^\alpha}
\Big[ \bar\psi(0) \hat z \psi(z)\Big]
\end{eqnarray}
where $z=(z^+,z^-,\vec{\bf z}_\perp)$.

Let us now consider the non-local operator $O^+(0,z)=\bar\psi(0) \gamma^+ \psi(z)$ where $z=z^\perp$ or $z=z^-$.
As above-mentioned, this operator has the definite collinear twist-$2$ (with a spin projection $s_a=+1$) and has no any definite geometrical twist.
On the other hand, after decomposition (see Eqn.~(\ref{non-loc})), the local operators which
form the given expansion can have the certain geometrical twists (and, therefore, the certain collinear twists):
\begin{eqnarray}
\label{loc-ops-gtw-1}
&&\bar\psi(0) \gamma^\alpha \psi(0) \quad \text{with} \quad \tau=2,
\end{eqnarray}
corresponding to the Lorentz spin $s=1$, and
\begin{eqnarray}
\label{loc-ops-gtw-2}
&&\bar\psi(0) \gamma^\alpha\vec{\cal D}^{\beta} \psi(0) \quad \text{leads to} \quad \tau=2,3,4
\end{eqnarray}
corresponding to the Lorentz spin $s=2,1,0$ and so on.
Indeed, one can write down that
\begin{eqnarray}
\label{tw-decom}
{\cal O}^{\alpha\beta}&=&\text{\Large\bf S}_{(\alpha\beta)}\,{\cal O}^{\alpha\beta} + {\cal O}^{[\alpha\beta]}
+ \frac{1}{4} g^{\alpha\beta} g_{\mu\nu} {\cal O}^{\mu\nu}
\nonumber\\
&\equiv& {\cal O}_{\{\tau=2\}} + {\cal O}_{\{\tau=3\}} +{\cal O}_{\{\tau=4\}}.
\end{eqnarray}
In a similar way we can treat any Lorentz high-rank tensors.
Moreover, to explain how the geometrical and collinear twists work, we consider the first two terms in the decompositions of
$O^+(0,z^\perp)$ and $O^+(0,z^-)$. We have
\begin{eqnarray}
\label{dec-nonloc-1}
&&\underbrace{\bar\psi(0) \gamma^+ \psi(z^\perp)}_{t=2, \, \text{no $\tau$-tw.}} =
\underbrace{\bar\psi(0) \gamma^+ \psi(0)}_{t=2 \,\,\text{from $\tau=2$}}
+ z^\perp \underbrace{\bar\psi(0) \gamma^+ \vec{\cal D}^{\perp} \psi(0)}_{t=3 \,\,\text{from $\tau=2,3$}} + ...
\nonumber\\
&&\underbrace{\bar\psi(0) \gamma^+ \psi(z^-)}_{t=2, \, \text{no $\tau$-tw.}} =
\underbrace{\bar\psi(0) \gamma^+ \psi(0)}_{t=2 \,\,\text{from $\tau=2$}}
+ z^- \underbrace{\bar\psi(0) \gamma^+ \vec{\cal D}^{+} \psi(0)}_{t=2 \,\,\text{from $\tau=2$}} + ...
\nonumber\\
\end{eqnarray}
Here, in the second terms the local operators of $t=3$ and $t=2$ receive the contributions from the local operators
of $\tau=2,3$ and $\tau=2$, respectively (see Eqn.~(\ref{tw-decom})). From eqn.~(\ref{dec-nonloc-1}), it has been seen
that the collinear twist of {\it l.h.s} and  {\it r.h.s}
conserves if we take into account the collinear twist of $z^\perp$ with $t=-1$ and $z^-$ with $t=0$.

\end{document}